\documentclass[aip,jcp,11pt,reprint]{revtex4-1}
\usepackage{booktabs}
\usepackage{graphicx}
\usepackage{color}%

\begin{document}

\title{Comment on ``Numerical Evidence Falsifying Finite-Temperature Many-Body Perturbation Theory"}
\author{Alec F. White}
\author{Garnet Kin-Lic Chan}
\affiliation{Division of Chemistry and Chemical Engineering, California Institute of Technology, Pasadena,
California 91125, USA}

\date{\today}

\begin{abstract}
In this comment we address the preprint of Jha and Hirata (arXiv:1809.10316 [physics.chem-ph])  which claims ``Numerical Evidence Falsifying Finite-Temperature Many-Body Perturbation Theory." We agree that finite difference differentiation of the exact grand potential is the correct way to verify the terms in the perturbation expansion. However, it is our suspicion that theoretical errors, uncontrolled numerical errors, and/or errors in their implementation have led Jha and Hirata to incorrectly conclude that finite temperature perturbation theory is incorrect. We show numerical evidence from finite difference differentiation of the grand potential that supports the correctness of finite-temperature perturbation theory. 
\end{abstract}
\maketitle

\section{Finite-temperature many-body perturbation theory}
A description of finite temperature perturbation theory can be found in a variety of textbooks,\cite{Blaizot1985,Negele1987,Fetter2003} and Santra and Schirmer have recently published a clear and pedagogical non-diagrammatic derivation.\cite{Santra2016} Jha's and Hirata's recent preprint\cite{Jha2018} expresses some uncertainty about the specifics of finite temperature perturbation theory, so we will first review the theory.

In imaginary time, the perturbation correction at $n$th order is proportional to $n$ nested integrals of the form
\begin{equation}
	\frac{1}{\beta} \int_0^{\beta}d\tau \int_0^{\tau} d\tau' \ldots \mathcal{T}
	\left[ V(\tau)V(\tau')\ldots \right]
\end{equation}
where $\mathcal{T}$ indicates an imaginary time-ordered product and $V(\tau)$ is the potential in the interaction picture. Note that for finite $\beta$, this expression consists of integrals of non-singular integrands over a finite region and is therefore {\it always finite for finite $\beta$}. This seems to contradict Jha's and Hirata's claim that conventional finite perturbation theory can yield infinite results at finite temperature. This misconception is perhaps due to statements in textbooks to the effect that
\begin{equation}\label{eqn:pt2_textbook}
	\Omega^{(2)} = \frac{1}{4}\sum_{ijab}\frac{|\langle ij|| ab\rangle|^2n_in_j
	(1 - n_a)(1 - n_b)}{\varepsilon_i + \varepsilon_j - 
	\varepsilon_a - \varepsilon_b}
\end{equation}
where the sum runs over all orbitals.
While this expression appears divergent, that is because it only applicable in the case that the divergent denominators are canceled by a zero interaction. This is often not clearly stated in textbooks, so some confusion is understandable. In the more general case, one finds
\begin{widetext}
\begin{eqnarray}\label{eqn:Omega2}
	\Omega^{(2)} &=& \frac{1}{4\beta}\sum_{ijab}|\langle ij|| ab\rangle|^2n_in_j
	(1 - n_a)(1 - n_b)\left[
	\frac{\beta}{\varepsilon_i + \varepsilon_j - \varepsilon_a - \varepsilon_b} + 
	\frac{1 - e^{\beta(
	\varepsilon_i + \varepsilon_j - \varepsilon_a - \varepsilon_b)}}
	{(\varepsilon_i + \varepsilon_j - \varepsilon_a - \varepsilon_b)^2}\right]
\end{eqnarray}
\end{widetext}
which is indeed always finite. If the perturbation includes a 1-body interaction, then there is a similar expression for the corresponding contribution from the singles. See Ref.~\onlinecite{Santra2016} for a more detailed derivation. From this expression, one can easily derive rules of the form given in Equations (11) and (15) of Jha's and Hirata's preprint.\cite{Jha2018} These expressions are called "unjustifiable" by Jha and Hirata, but they arise naturally from the derivation of finite temperature perturbation theory.

Finally, we must mention that these expressions are all rigorously derived as corrections to the grand potential. Corrections to the energy can also be derived, but they include some additional terms. Santra and Schirmer discuss this in detail.\cite{Santra2016} Jha and Hirata claim that Mattuck\cite{Mattuck1992} and Fetter and Walecka\cite{Fetter2003} both assign corrections of the type given in Equation~\ref{eqn:pt2_textbook} to the internal energy, but we cannot find any statements to that effect in Fetter and Walecka. Mattuck contains an equation which implies a diagrammatic expansion of the internal energy, but the interpretation of this
diagrammatic equation is not elaborated on. Furthermore, Mattuck references ``Bloch (1962)''\cite{BLOCH1961241} which describes a perturbation expansion of the grand potential. We do not see this issue as a matter of much debate. 

\section{Test systems}
To test numerically the values of the corrections we will focus on two systems: Be atom and HF molecule both in minimal basis sets.

\subsection{Be atom, sto-3g}
For this test, we will use Be atom in a minimal basis. We use the orbitals and orbital energies computed from a zero temperature Hartree-Fock calculation on the neutral molecule. This choice of reference means that the perturbation will contain a 1-body part, but it makes the reference easily reproducible. For all calculations we set $\mu = 0$. This means that the number of particles in the reference systems will change at different temperatures, but this causes no problem because the grand canonical ensemble is defined for fixed $\mu$. 

\subsection{HF molecule, sto-3g}
We will also run calculations on the HF molecule in a minimal basis at a bond length of 0.9168 \AA . This is the same system as considered by Jha and Hirata.\cite{Jha2018} We set $\mu$ such that the reference systems is neutral as done by Jha and Hirata. However, we still use the zero temperature orbitals and orbital energies. Jha and Hirata use a different reference, but we were unable to reproduce their results; it is not entirely clear how they chose the reference orbitals. We use the zero temperature orbitals and energies as a reference. We should also note that this system is numerically challenging. At high temperatures there are so few orbitals, that a very large value of $\mu$ must be used to ensure the reference is neutral. We report the value of $\mu$ that is used at each temperature so that our calculations can be easily verified.

\section{Methods and formulas}

\subsection{Analytic formulas}
We will mimic the notation of Jha and Hirata and use $\tilde{\Omega}^{(n)}$ to indicate the $n$th order correction calculated from analytic formulas.
The 0th order part of the grand potential is given by
\begin{equation}
	\tilde{\Omega}^{(0)} = E_{\text{nuc}} + \frac{1}{\beta}\sum_i\ln n_i
\end{equation}
where the sum runs over all orbitals and $n_i$ is the Fermi occupation of the $i$th orbital. The first order correction is given by
\begin{equation}
	\tilde{\Omega}^{(1)} = \sum_i (f_{ii} - \varepsilon_i)n_i - \frac{1}{2}
	\sum_{ij} \langle ij || ij \rangle n_in_j.
\end{equation}
Here the sum again runs over all orbitals, $\varepsilon_i$ indicates the $i$th reference energy, and $f$ is the finite temperature Fock operator defined as
\begin{equation}
	f_{pq} \equiv h_{pq} + \sum_{j}\langle pj || qj \rangle n_j.
\end{equation}

At second order, we first define the 1-body part of the perturbation as
\begin{equation}
	\tilde{f}_{pq} \equiv f_{pq} - \delta_{pq}\varepsilon_p.
\end{equation}
The 2nd order correction is given by
\begin{widetext}
\begin{eqnarray}\label{eqn:Omega2}
	\tilde{\Omega}^{(2)} &=& \frac{1}{4\beta}\sum_{ijab}|\langle ij|| ab\rangle|^2n_in_j
	(1 - n_a)(1 - n_b)\left[
	\frac{\beta}{\varepsilon_i + \varepsilon_j - \varepsilon_a - \varepsilon_b} + 
	\frac{1 - e^{\beta(
	\varepsilon_i + \varepsilon_j - \varepsilon_a - \varepsilon_b)}}
	{(\varepsilon_i + \varepsilon_j - \varepsilon_a - \varepsilon_b)^2}\right]
	\nonumber \\
	&+&\frac{1}{\beta}\sum_{ia}|\tilde{f}_{ai}|^2n_i(1 - n_a)\left[
	\frac{\beta}{\varepsilon_i - \varepsilon_a} +
	\frac{1 - e^{\beta(\varepsilon_i - \varepsilon_a)}}
	{(\varepsilon_i - \varepsilon_a)^2}\right].
\end{eqnarray}
\end{widetext}
The sums run over all orbitals.

\subsection{The grand potential from exact diagonalization}
In order to numerically compute the $n$th order correction to the grand potential, we define the Hamiltonian as a function of the coupling as
\begin{equation}
	H(\lambda) = \varepsilon + \lambda(h - \varepsilon + V)
\end{equation}
where $\varepsilon$ is the diagaonal operator defining the non-interacting reference, $h$ is the core hamiltonian, and $V$ is the coulomb operator. We can then compute the $n$th order correction to $\Omega$ as
\begin{equation}
	\Omega^{(n)} = \frac{1}{n!} \left.\frac{d^n\Omega(\lambda)}{d\lambda^n}
	\right|_{\lambda = 0}.
\end{equation}

This requires that we compute $\Omega$ exactly as a function of $\lambda$. We accomplish this by computing the partition function directly. All full-CI states for all possible numbers of electrons are computed, and for each state, we compute the contributions to the grand partition function:
\begin{equation}
	Z_g = \sum_{\nu} e^{-\beta(E_{\nu} - \mu N_{\nu})}.
\end{equation} 
The grand potential is then computed as
\begin{equation}
	\Omega = -\frac{1}{\beta}\ln Z_g.
\end{equation}

This is difficult accomplish numerically because of the danger of overflow/underflow. To circumvent this issue, we compute the exponential using high-precision floats from the mpmath package.\cite{mpmath} These floats store the exponent exactly in binary representation and we used 100 decimal places of precision. In this way, we avoid underflow/overflow and the numerical instabilities of direct computation of the partition function so that the precision of the final answer is at worst limited by the precision of the FCI calculation itself (roughly machine precision for double precision floats).

\subsection{Finite difference differentiation}
\begin{figure}[h]
\includegraphics[scale=1.0]{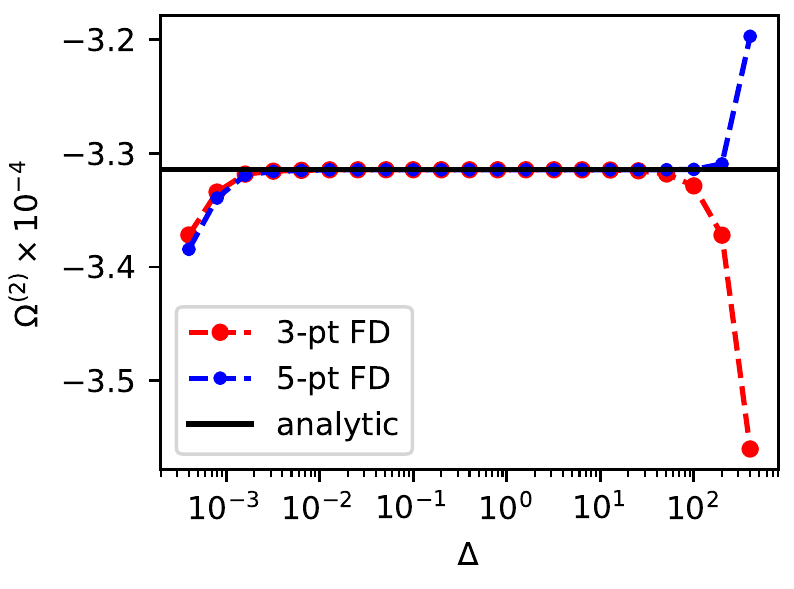}
\caption{\label{fig:fd} Finite difference computation of $\Omega^{(2)}$ of the HF molecule for different values of the difference, $\Delta$. The analytic result is shown for comparison. Notice that at high temperatures ($T = 1.0\text{E}9$ K for this case) large values of the difference may be necessary to ensure a reasonable accuracy.} 
\end{figure}

The 1st order and 2nd order contributions to the grand potential are computed by finite differences as
\begin{equation}
	\Omega^{(1)} = \left.\frac{d\Omega(\lambda)}{d\lambda}\right|_{\lambda = 0}
	\qquad \Omega^{(2)} = \frac{1}{2}
	\left.\frac{d^2\Omega(\lambda)}{d\lambda^2}\right|_{\lambda = 0}.
\end{equation}
where we evaluate the derivatives using a 3-point or 5-point stencil. A good value of the finite difference, $\Delta$, can be determined by plotting the derivative for a wide range of $\Delta$, and identifying the plateau region. For the first derivatives this is hardly necessary, but for second derivatives at high temperatures, the second derivative can be 10 orders of magnitude smaller than the value of the function and much larger values of $\Delta$ are required. As an example of this behavior, see Figure~\ref{fig:fd}. 

\section{Results}
In this section we show that the 1st and 2nd order corrections computed from the analytic expressions are consistent with the results obtained from finite difference differentiation. We have made the code that computes the PT publicly available.\footnote{Code publicly available: https://github.com/awhite862/ft\_mp2\_test}

\subsection{Be atom, sto-3G}
The results for Be atom are shown in Table~\ref{tab:Be}.
\begin{table*}[h]
\begin{ruledtabular}
\begin{tabular}{ccccccc}
  $T$ (K) & $\mu$  & $\Delta$ & $\tilde{\Omega}^{(1)}$ &$\Omega^{(1)}$ &error &Rel. error \\
  \hline
  1.0E3& 0.0&	1.0E-4& -4.875820876&	-4.875820876&	1.7E-10&	-3.4E-11\\
  1.0E4& 0.0&	1.0E-4& -4.875816983&	-4.875816983&	1.7E-10&	-3.4E-11\\
  1.0E5& 0.0&	1.0E-4& -4.624519973&	-4.624519956&	1.7E-08&	-3.6E-09\\
  1.0E6& 0.0&	1.0E-4& -4.740836562&	-4.740836562&	3.5E-10&	-7.3E-11\\
  1.0E7& 0.0&	4.0E-4& -4.931617917&	-4.931617917&	2.5E-10&	-5.1E-11\\
  1.0E8& 0.0&	3.2E-3& -4.934561248&	-4.934561248&	8.6E-11&	-1.7E-11\\
  1.0E9& 0.0&	2.6E-2& -4.934625218&	-4.934625218&	3.4E-11&	-6.9E-12
 \end{tabular}
\end{ruledtabular}
\begin{ruledtabular}
\begin{tabular}{ccccccc}
  $T$ (K) & $\mu$  & $\Delta$ &  $\tilde{\Omega}^{(2)}$ &$\Omega^{(2)}$ &error &Rel. error \\
  \hline
  1.0E3& 0.0&	1.00E-04&  -0.024358375&	-0.024358293& 8.1E-08& 3.3E-06 \\
  1.0E4& 0.0&	1.00E-04&  -0.024339086&	-0.024339020& 6.6E-08& 2.7E-06 \\
  1.0E5& 0.0&	1.00E-04&  -0.985577556&	-0.985577575& 1.9E-08& 2.0E-08 \\
  1.0E6& 0.0&	1.00E-04&  -0.187353714&	-0.187354310& 6.0E-07& 3.2E-06 \\
  1.0E7& 0.0&	4.00E-04&  -0.016083535&	-0.016083557& 2.2E-08& 1.3E-06 \\
  1.0E8& 0.0&	3.20E-03&  -0.001579361&	-0.001579292& 6.9E-08& 4.4E-05 \\
  1.0E9& 0.0&	2.56E-02&  -0.000157658&	-0.000157618& 3.9E-08& 2.5E-04
 \end{tabular}
\end{ruledtabular}
\caption{\label{tab:Be} Numerical vs. analytic computation of the first and second order corrections to the grand potential for Be atom.}
\end{table*}

\subsection{HF molecule, sto-3G}
The results for HF molecule are shown in Table~\ref{tab:HF}. 
\begin{table*}[h]
\begin{ruledtabular}
\begin{tabular}{ccccccc}
  $T$ (K) & $\mu$  & $\Delta$ & $\tilde{\Omega}^{(1)}$ &$\Omega^{(1)}$ &error &Rel. error \\
  \hline
1.0E3&  0.000000E+00&	8.00E-04& -45.99585606& -45.99585606& 3.4E-11& 7.4E-13\\
1.0E4&  9.368177E-02&	3.20E-03& -45.99585606& -45.99585606& 6.2E-12& 1.3E-13\\
1.0E5&  2.722387E-01&	3.20E-03& -46.02031821& -46.02031821& 9.9E-10& 2.1E-11\\
1.0E6&  3.961354E+00&	3.20E-03& -46.21519215& -46.21519215& 4.2E-10& 9.0E-12\\
1.0E7&  4.715075E+01&	3.20E-03& -46.18019361& -46.18019361& 1.7E-10& 3.6E-12\\
1.0E8&  5.050709E+02&	6.40E-03& -46.10674651& -46.10674651& 4.1E-11& 8.8E-13\\
1.0E9&  5.092157E+03&	1.28E-02& -46.09626049& -46.09626049& 7.1E-11& 1.5E-12
  \end{tabular}
\end{ruledtabular}
\begin{ruledtabular}
\begin{tabular}{ccccccc}
  $T$ (K) & $\mu$  & $\Delta$ &  $\tilde{\Omega}^{(2)}$ &$\Omega^{(2)}$ &error &Rel. error \\
  \hline
1.0E3& 0.000000E+00& 8.00E-04& -0.017335597& -0.017335654& 5.7E-08&	3.3E-06\\
1.0E4& 9.368177E-02& 3.20E-03& -0.017335602& -0.017335599& 2.9E-09&	1.7E-07\\
1.0E5& 2.722387E-01& 3.20E-03& -0.268951310& -0.268951296& 1.4E-08&	5.1E-08\\
1.0E6& 3.961354E+00& 3.20E-03& -0.120552925& -0.120552923& 2.1E-09&	1.8E-08\\
1.0E7& 4.715075E+01& 3.20E-03& -0.021837989& -0.021837988& 8.9E-10&	4.1E-08\\
1.0E8& 5.050709E+02& 6.40E-03& -0.003181121& -0.003181118& 3.0E-09&	9.6E-07\\
1.0E9& 5.092157E+03& 1.28E-02& -0.000331449& -0.000331456& 7.0E-09&	2.1E-05
   \end{tabular}
\end{ruledtabular}
\caption{\label{tab:HF} Numerical vs. analytic computation of the first and second order corrections to the grand potential for HF molecule.}
\end{table*}

\section{Conclusions}
We have presented strong numerical evidence to confirm the correctness of traditional finite-temperature perturbation theory. As this theory has been used for many decades without issue, this is the expected result. We have not been able to perform precisely the same calculations as Hirata and Jha, because we are not sure of the specifics of their reference. However, we see no reason why a different choice of reference would lead to dramatically different results.

\begin{acknowledgements}
This work is supported by the US Department of Energy, Office of Science, via grant number SC0018140.
\end{acknowledgements}

\clearpage
\section{References}
\bibliography{ref}

\end{document}